# Financial black swans driven by ultrafast machine ecology

Neil Johnson[1], Guannan Zhao[1], Eric Hunsader[2], Jing Meng[1], Amith Ravindar[1], Spencer Carran[1] and Brian Tivnan[3,4]

[1] Physics Department, University of Miami, Coral Gables, Florida 33124, U.S.A.

[2] Nanex LLC, Evanston, Illinois, U.S.A.

[3] The MITRE Corporation, McLean, VA 22102, U.S.A.

[4] Complex Systems Center, University of Vermont, Burlington, VT 05405, U.S.A.

**ABSTRACT**

Society's drive toward ever faster socio-technical systems[1-3], means that there is an urgent need to understand the threat from 'black swan' extreme events that might emerge[4-19]. On 6 May 2010, it took just five minutes for a spontaneous mix of human and machine interactions in the global trading cyberspace to generate an unprecedented system-wide Flash Crash[4]. However, little is known about what lies ahead in the crucial sub-second regime where humans become unable to respond or intervene sufficiently quickly[20,21]. Here we analyze a set of 18,520 ultrafast black swan events that we have uncovered in stock-price movements between 2006 and 2011. We provide empirical evidence for, and an accompanying theory of, an abrupt system-wide transition from a mixed human-machine phase to a new all-machine phase characterized by frequent black swan events with ultrafast durations (<650ms for crashes, <950ms for spikes). Our theory quantifies the systemic fluctuations in these two distinct phases in terms of the diversity of the system's internal ecology and the amount of global information being processed. Our finding that the ten most susceptible entities are major international banks, hints at a hidden relationship between these ultrafast 'fractures' and the slow 'breaking' of the global financial system post-2006. More generally, our work provides tools to help predict and mitigate the systemic risk developing in any complex socio-technical system that attempts to operate at, or beyond, the limits of human response times.

The downside of society's continuing drive toward larger, faster, and more interconnected socio-technical systems such as global financial markets[1,3], is that future catastrophes may be less easy to forsee and manage -- as witnessed by the recent emergence of financial flash-crashes[1,3,4]. In traditional human-machine systems, real-time human intervention may be possible if the undesired changes occur within typical human reaction times. However, even if one sets aside the time taken for humans to physically react, it takes a chess grandmaster approximately 650 milliseconds just to realize that she is in trouble (i.e. her king is in checkmate)[20,21]. In many areas of human activity, the quickest that someone can notice such a cue and physically react, is approximately 1000 milliseconds (1 second)[20,21]. Notwithstanding this biophysical limitation, the strategic advantage to a financial company of having a faster system than its competitors is currently driving a billion-dollar technological arms race[22,23] to reduce communication and computational operating times down toward the physical limits of the speed of light – orders of magnitude below human response times[22,23]. For example, a new dedicated transatlantic cable[22] is being built just to shave 5 milliseconds off transatlantic communication times between US and UK traders, while a new purpose-built chip iX-eCute is being launched which prepares trades in 740 nanoseconds (1 nanosecond is $10^{-9}$ seconds)[23]. Financial market price movements are determined by the self-organized activity of a global collective of trading agents[1-19], including both humans and machine algorithms. Like many other complex systems, financial markets have no real-time controller, hence mitigation relies on carefully choosing the trading regulations. However, as stressed by the seminal works of Farmer, Preis, Stanley, Cliff and co-workers[1,8,19], effective regulation will remain elusive until researchers develop a deep quantitative understanding of the system's dynamics, and hence a scientific theory for the underlying human-machine ecology on these ultrafast timescales[1,8,14,19,24].

Motivated by these parallel scientific and practical goals, and the fact that financial markets are the world's largest and most data-rich example of an ultrafast, self-organizing socio-technical system, we undertook a search for ultrafast extreme events in a high-throughput millisecond-resolution stream of prices for multiple stocks across multiple exchanges between 2006-2011. (For convenience, we use the popular term[5] 'black swan' for each extreme event while mindful of Sornette's important 'dragon king' terminology[6]). For a large price drop to qualify as an extreme event (i.e. black swan crash) the stock price had to tick down at least ten times before ticking up and the price change had to exceed 0.8%. For a large price rise to qualify as an extreme event (i.e. black swan spike) the stock had to tick up at least ten times before ticking down and the price change had to exceed 0.8%. In order to explore timescales which go beyond typical human reaction times[20,21], we focus on black swans with durations less than 1500 milliseconds. We uncovered 18,520 such black swan events, which surprisingly is more than one per trading day on average. As stressed in Ref. [7] following the 16th century philosopher Francis Bacon, the scientific appeal of extreme events is that it is in such moments that a complex system offers glimpses into the true nature of the underlying fundamental forces that drive it[7].

Figure 1 illustrates a crash (Fig. 1A) and spike (Fig. 1B) from our dataset, both with duration 25 milliseconds (0.025s), while Fig. 1C suggests a systemic coupling between these sub-second black swan events in individual stock (blue and red curves) and long-term market-wide instability on the scale of weeks, months and even years (black curve). Each black swan feature in Figs. 1A and 1B is huge compared to the size of the fluctuations either immediately before or after it, while the quick recovery from the initial drop or rise probably results from an automatically triggered exchange response or predatory computer trades[3]. The coupling in Fig. 1C across such vastly different timescales is made even more intriguing by the fact that the ten stock with highest incidences of ultrafast black swans are all financial institutions -- and yet it is financial institutions that have been most strongly connected with the late 2000's global financial collapse (e.g. Lehmann Brothers filing for Chapter 11 bankruptcy protection on 15 September 2008). This suggests an analogy to engineering systems where it is well-known that a prevalence of micro-fractures can accompany, and even precede, large changes in a mechanical structure (e.g. tiny cracks in a piece of plane fuselage which then eventually breaks off). Indeed, our dataset shows a far greater tendency for these financial fractures to occur, within a given duration time-window, as we move to smaller timescales, e.g. 100-200ms has approximately ten times more than 900-1000ms. The fact that the instantaneous rate of occurrence of spikes and crashes is similar (i.e. blue and red curves are almost identical in Fig. 1C) suggests that these ultrafast black swans are not simply the product of some pathological regulatory rule for crashes, for example the uptick rule[3,4]. An immediate implication of these observations for regulators is that extreme behaviors on very short (i.e. < 1s) and long timescales (e.g. 1 year, or $\sim 10^{-7}$s) cannot *a priori* be separated, hence rules targeted solely at controlling intraday fluctuations, or calming markets on the scale of months or years, can induce dangerous feedback effects at the opposite timescale.

Figure 2 shows that the character of the ultrafast black swans changes fundamentally as the duration threshold is reduced beyond typical human reaction times[20,21]. The statistical tool used is the power-law -- specifically the best-fit power-law exponent $\alpha$ and goodness-of-fit *p* value (see Methods Summary) for the size distribution of black swans in our dataset with duration equal to or greater than a threshold value $\tau$. The choice of power-law is motivated by the fact that price movements are known to exhibit power-law statistics across a range of longer timescales[6-13]. Other statistical measures also show this transition behavior, hence the result in Fig. 2 is not a spurious effect of our analysis. For black swans of duration $\tau > 1000$ms (i.e. 1 second) the power-law distribution is strongly supported with *p* values near 1. By contrast, the inclusion of black swans with ever smaller durations rapidly changes the distribution away from a power-law. The transitions in Fig. 2 occur over a remarkably short timescale and range, and finish by 1000ms which is a typical timescale for the decision-plus-action of a human who is neither the highest-trained expert nor is 100% attentive (e.g. a typical motorist on a typical day[20,21]).

A chess grandmaster takes approximately 650 milliseconds to realize that her king is in checkmate[20,21] -- hence even if a human trader is as attentive and capable as a chess grandmaster, it would take him 650 milliseconds to realize he is in danger of losing (i.e. king is in checkmate[20,21]) before actually making a move to buy or sell. Given that a market is a collection of autonomous human and machine (i.e. computer algorithm) agents watching the latest prices before their next move, and yet human beings have limitations on how fast they can notice a particular situation and act on it, it is reasonable to interpret the transitions in Fig. 2 as reflecting the decreasing ability of human beings to influence price movements at smaller timescales. It is remarkable that the collective action of the global trading population mimics the timescale of a chess grandmaster's thought process by kicking in at 650ms for crashes. This can perhaps be explained by noting that any unexpected sharp downward turn in the price would be seen as dangerous by the majority of market participants who are typically 'long' the market[3], and hence will be acted upon as quickly as physically possible. By contrast, a sharp upward movement might typically be regarded as positive and hence require no rapid action. This fast collective action for crashes is also reminiscent of the 'many eyes' idea from ecology that describes why large groups rapidly detect imminent danger, and hence supports Farmer and Skouras' fascinating ecological perspective[1].

Our proposed theory associates these findings with a new fundamental transition from a mixed phase of humans and machines, in which humans have time to assess information and act, to an ultrafast all-machine phase in which machines dictate price changes. Our model considers an ecology of $N$ heterogenous agents (machines and/or humans) who repeatedly compete to win in a competition for limited resources[25-30]. Each agent possesses $s > 1$ strategies[26-30]. An agent only participates if it has a strategy that has performed sufficiently well in the recent past[27,28]. It uses its best strategy at a given timestep[26-30]. The agents sit watching a common source of information, e.g. recent price movements[26-30] encoded as a bit-string of length $m$, and act on potentially profitable patterns they observe. Consider the regime in which the total number of different strategies in the market (which is $2^{m+1}$ in our model) is typically larger than the total number of agents $N$ (i.e. $\eta > 1$ where $\eta = 2^{m+1}/N$). We associate this regime (see Fig. 3) with a market in which both humans and machines are dictating prices, and hence timescales above the transition (>1s), for these reasons: The presence of humans actively trading -- and hence their 'free will' together with the myriad ways in which they can manually override algorithms -- means that the effective number (i.e. diversity) of strategies should be large (i.e. $\eta > 1$). Moreover $\eta > 1$ implies $m$ is large, hence there are more pieces of information available which suggests longer timescales (there will be more millisecond price movements in the past 1000ms than in the past 500ms). Since by definition $N/2^{m+1} < 1$ in this $\eta > 1$ regime, the average number of agents per strategy is less than 1, hence any crowding effects due to agents coincidentally using the same strategy will be small. This lack of crowding leads our model to predict that any large price movements arising for $\eta > 1$ will be rare and take place over a longer duration – exactly as observed in our data for timescales above 1000ms. Indeed, our model's price output (e.g. Fig. 3, right-hand panel) reproduces the stylized facts associated with financial markets over longer timescales, including a power-law distribution[5-19] (see Ref. [27]). Our model undergoes a transition around $\eta \approx 1$ to a regime characterized by significant strategy crowding (see Ref. [27] for full details) and hence large fluctuations. The price output for $\eta < 1$ (Fig. 3, left-hand panel) shows frequent abrupt changes due to agents moving as unintentional groups into particular strategies. Our model therefore predicts a rapidly increasing number of ultrafast black swan events as we move to smaller $\eta$ and hence smaller subsecond timescales – as observed in our data. Our association of the $\eta < 1$ regime with an all-machine phase is consistent with the fact that trading algorithms in the sub-second regime need to be executable extremely quickly and hence be relatively simple, without calling on much memory concerning past information: Hence $m$ will be small, so the total number of strategies will be small and therefore $2^{m+1} < N$ which means $\eta < 1$. Our model also predicts that the size distribution for the black swans in this ultrafast regime ($\eta < 1$) should not have a power law since changes of all sizes do not appear – this is again consistent with the results in Fig. 2.

Although our model ignores many potentially important details about the real market, its simplicity allows us to derive analytic formulae for the scale of the price fluctuations in each phase if we make the additional assumption that the number of agents playing each timestep is similar to $N$. For $\eta < 1$, the standard deviation $\sigma$ of the price fluctuations has a lower bound given by (see Ref. [27]) $\sigma = 3^{-\frac{1}{2}} 2^{-\frac{(m+1)}{2}} N \left(1 - 2^{-2(m+1)}\right)^{\frac{1}{2}}$ for $s = 2$, and an upper bound which is a factor of $\sqrt{2}$ bigger given by $\sigma = 3^{-\frac{1}{2}} 2^{-\frac{m}{2}} N \left(1 - 2^{-2(m+1)}\right)^{\frac{1}{2}}$. For $\eta > 1$, $\sigma$ is given approximately by $\sigma = N^{\frac{1}{2}} \left(1 - 2^{-(m+1)} N\right)^{\frac{1}{2}}$ for general $s$. Our model's prediction that $\sigma$ is proportional to $N$ for $\eta < 1$ as compared to $N^{\frac{1}{2}}$ for $\eta > 1$, provides an analytic explanation for the empirical finding that there are many more black swans at shorter durations, while the transition in $\sigma$ around $\eta \approx 1$ explains the abrupt change in the character of their distribution observed in Fig. 2. Since $\sigma$ plays a fundamental role in traditional finance as a measure of risk, these explicit formulae and their parameter dependencies could be used to help quantify the effect of changes in regulations on conventional risk measures.

Figure 4A confirms that recoveries can emerge spontaneously from our model (in particular for the $\eta < 1$ regime corresponding to very short durations as in Figs. 1A and 1B) without having to invoke external regulations or additional predatory algorithms[3], and guides us toward a specific prediction and mitigation scheme. The nodes in Fig. 4B show the possible global information (i.e. price-history) bit-strings $\mu$ for the case $m = 3$, while the possible transitions giving negative (positive) price-changes and hence producing an update of 0 (1) to $\mu$, are shown as red (blue) arrows. As the model updates $\mu$ at each timestep, it traces a trajectory around the network in Fig. 4B. At any given time, the strategy score vector can be expanded as $\underline{S}[t] = \sum_{j=0}^{2^m-1} c_j \underline{a}^j$ where the vectors $\underline{a}^j$ are binary basis vectors and the expansion coefficients $c_j$ represent effective weights for each node $j$. Each node acts like a coiled spring, in that the bigger $c_j$, the greater the tendency of the system to return to that node. Any trajectory that comprises mostly negative (positive) transitions will produce a large price drop (rise). Figure 4A shows these weightings and the overall model trajectory (green line) as it moves between nodes, expressed in their decimal representation with $\mu = \{000\}$ equivalent to 0 etc. Prior to the initial price drop in Fig. 4A, there is a large positive weight (blue) on node 0 (i.e. at $\mu = \{000\}$ in Fig. 4B). When the model's trajectory hits node 0 on timestep 346, this large weight triggers repeated transitions back to node 0, like a spring uncoiling, producing a large number of consecutive negative price changes -- hence the large price drop in Fig. 4A. Node 7 (i.e. $\mu = \{111\}$ in Fig. 4B) has a large weight of the opposite sign, which means that when it is subsequently hit, the reverse happens and the price rebounds as shown in Fig. 4A, consistent with the observation in Fig. 1A (and 1B by symmetry). Reference [27] shows analytically that the duration of the drop, and by symmetry the rebound, is $c_0$ timesteps. Each model timestep represents the minimum update time for information in the market[3] (e.g. 2ms). Future work will test if it is possible to accurately predict the total duration of real market mini-crashes prior to them unfolding, simply by counting prior outcomes in order to estimate the current nodal weight $c_0$. If the weight on node 7 is not large, there will be no recovery and the rapid drop will show no rebound, as in Fig. 3 left panel. This analysis assumes no exogenous kicks from external news -- however this assumption becomes increasingly justifiable as the ultrafast black swan duration decreases and hence the possibility of major external news arriving, being assimilated, and acted upon, becomes less likely[3].

Figure 4C shows an intervention scheme for mitigating these subsecond black swan events. Consistent with the idea of Satinover and Sornette[28], the strong weights which emerge on the extreme nodes 0 and 7 prior to a large change provide an inherent predictability which helps cancel the stochasticity generated by 'undecided' agents (i.e. agents holding several tied highest-scoring strategies which each suggest opposite actions)[27-30]. Just prior to a large change, the resulting system is therefore (momentarily) largely deterministic. If left alone, it will produce a large change in a definite direction – however, it can be driven away from the impending crash or spike by increasing the strategy diversity in the following ways, listed here in order of increasing effect but also increasing assumed knowledge: (1) If little is known about the system, our earlier expressions for $\eta < 1$ predict that increasing the disorder in the initial strategy allocations by adding agents with randomly chosen strategies will reduce the scale of the fluctuations toward their lower bound, and hence reduce the size of the black swan by a factor of up to $\sqrt{2}$. (2) If something is known about the most popular strategies currently in use, inserting new agents with strategies that are anti-correlated to these popular strategies, should reduce the black swan sizes by an order of magnitude[27]. (3) If the overall occupation of different strategies is known as in Fig. 4C (blue boxes) then even without knowing the actual strategies for individual agents, placing just a few additional agents at specific locations in the occupancy matrix (red boxes) can produce a trajectory that completely bypasses the black swan[29]. The next stage for this work would be to identify specific remedies for specific black-swan scenarios.

It will be interesting to see if other explanations for our data, and hence Fig. 2, can be established. Distinguishing between such explanations may require additional data concerning the identity of trades together with specific exchange details at the millisecond scale. However, the generality of our model suggests that the transition we identify between a mixed human-machine and all-machine ecology could arise in a variety of socio-technical systems operating near the limits of human response times, and opens up the possibility of a common regulatory approach for tackling ultrafast black swan events across a wider class of systems.

We gratefully acknowledge support for this research from The MITRE Corporation. The views and conclusions contained in this paper are those of the authors and should not be interpreted as representing the official policies, either expressed or implied, of any of the above named organizations, to include the U.S. government. NFJ thanks Zhenyuan Zhao, Pak Ming Hui, David Smith, Michael Hart and Paul Jefferies for discussions surrounding this topic.

**Figure Captions**

**Figure 1:** Traded price during black swan events. (A) Crash. Stock symbol is ABK. Date is 11/04/2009. Number of sequential down ticks is 20. Price change is -0.22. Duration is 25ms (i.e. 0.025 seconds). Percentage price change downwards is 14% (i.e. crash magnitude is 14%). (B) Spike. Stock symbol is SMCI. Date is 10/01/2010. Number of sequential up ticks is 31. Price change is +2.75. Duration is 25ms (i.e. 0.025 seconds). Percentage price change upwards is 26% (i.e. spike magnitude is 26%). Dots in price chart are sized according to size of trade. (C)

Cumulative number of crashes (red) and spikes (blue) compared to overall stock market index (Standard & Poor's 500) in black, showing daily close data from 3 Jan 2006 until 3 Feb 2011.

**Figure 2:** Empirical transition in size distribution for black swans with duration above threshold $\tau$, as function of $\tau$. Top: Scale of times. 650 ms is the time for chess grandmaster to discern King is in checkmate. Plots show the results of the best-fit power-law exponent (black) and goodness-of-fit (blue) to the distributions for size of crashes and spikes separately, as shown in the inset schematic.

**Figure 3:** Theoretical transition. Model output for the two regimes of strategy distribution among agents ($\eta = 2^{m+1}/N$) together with timescales from Fig. 2 (top). $\eta < 1$ implies many agents per strategy, hence large crowding which produces frequent, large and abrupt price-changes, i.e. high number of short-duration ($\ll 1$ second) black swans, as observed empirically. $\eta > 1$ implies very few, if any, agents per strategy, hence small crowding. Therefore large changes are rarer and last longer, i.e. low number of longer-duration black swans, as observed empirically.

**Figure 4:** Prediction and mitigation of ultrafast black swans. (A) Theoretical black swan produced by our model, similar to Fig. 1A on an expanded timescale. Below are model's node weights for $m = 3$ as a function of time, with blue and red denoting the weight values (see text and SI). The more blue the weight on node 0 (or the more red the weight on node 7) the more likely that a large price-drop (or rise) will occur when the model's trajectory (green curve) hits that node. (B) Nodes shown as their binary and decimal equivalents (e.g. 000 is 0 in decimal) for $m = 3$. Red (blue) arrows represent transitions generating a bit-string update of 0 (1) and hence a price drop (rise). (C) Proposed mitigation scheme. Blue boxes show histogram of strategy possession for existing population, as an 'occupancy' in a two-dimensional space spanned by the possible strategy

combinations for $s = 2$ strategies with $m = 2$. Red boxes are inserted agents, which can induce a steering effect to avoid the crash[29] (red dotted line in Fig. 4A showing the mean of the resulting price movement). See text for discussion.

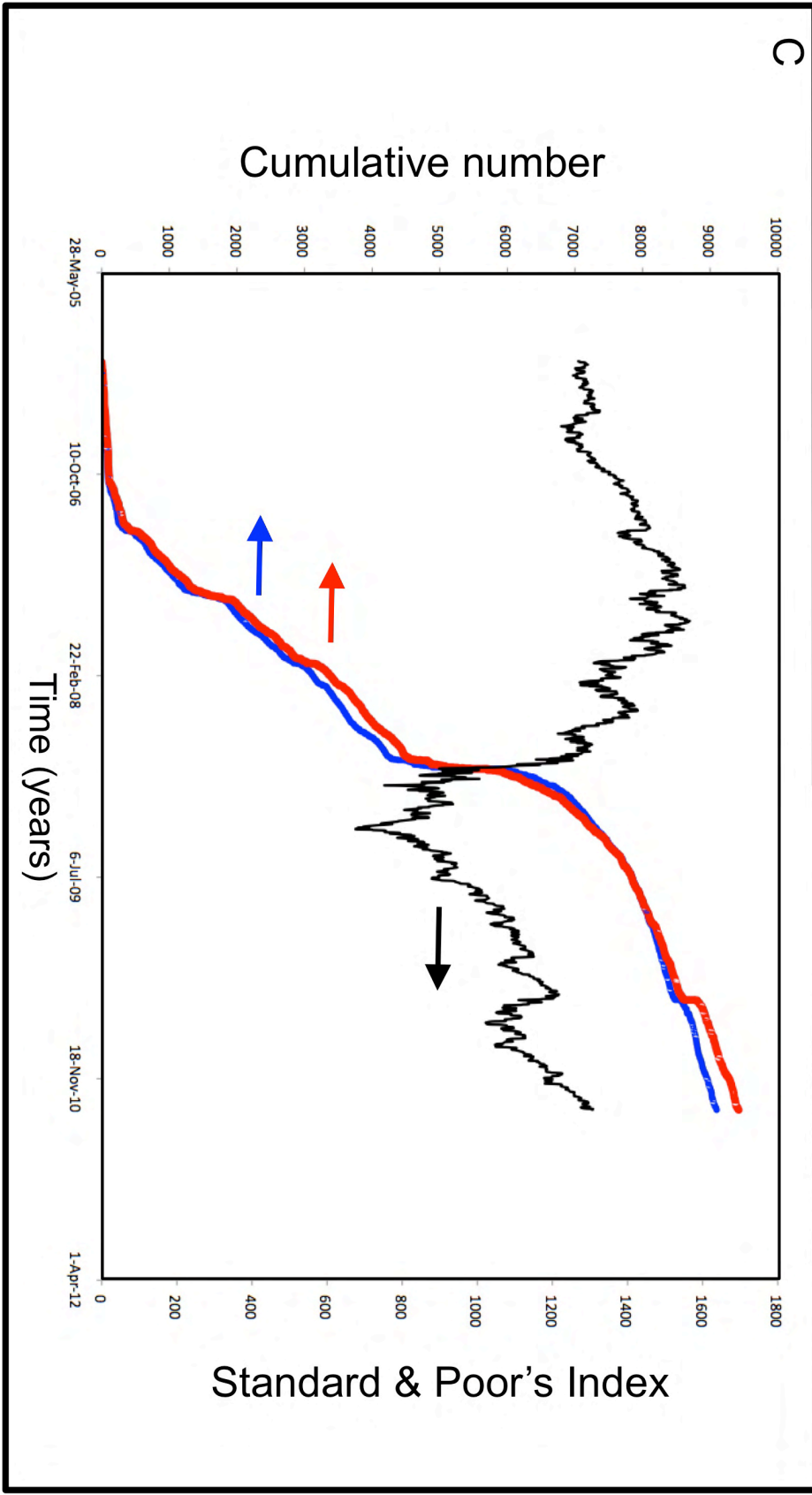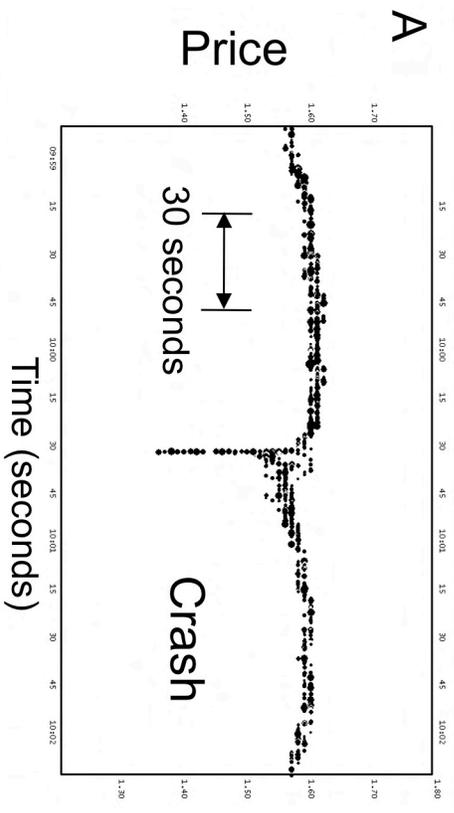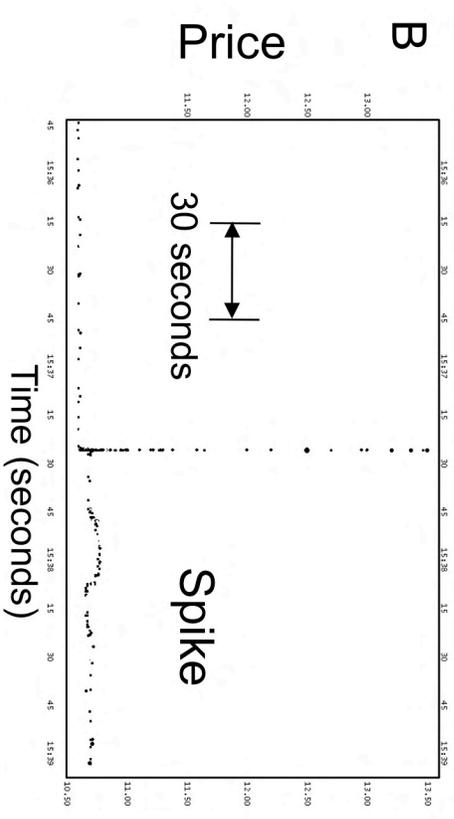

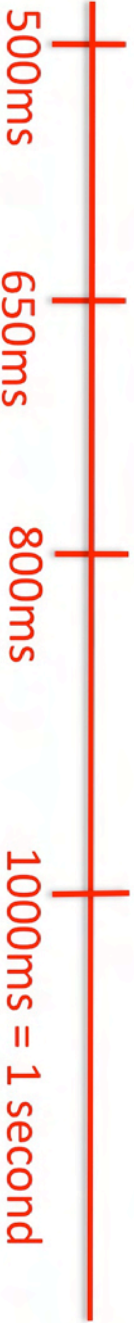
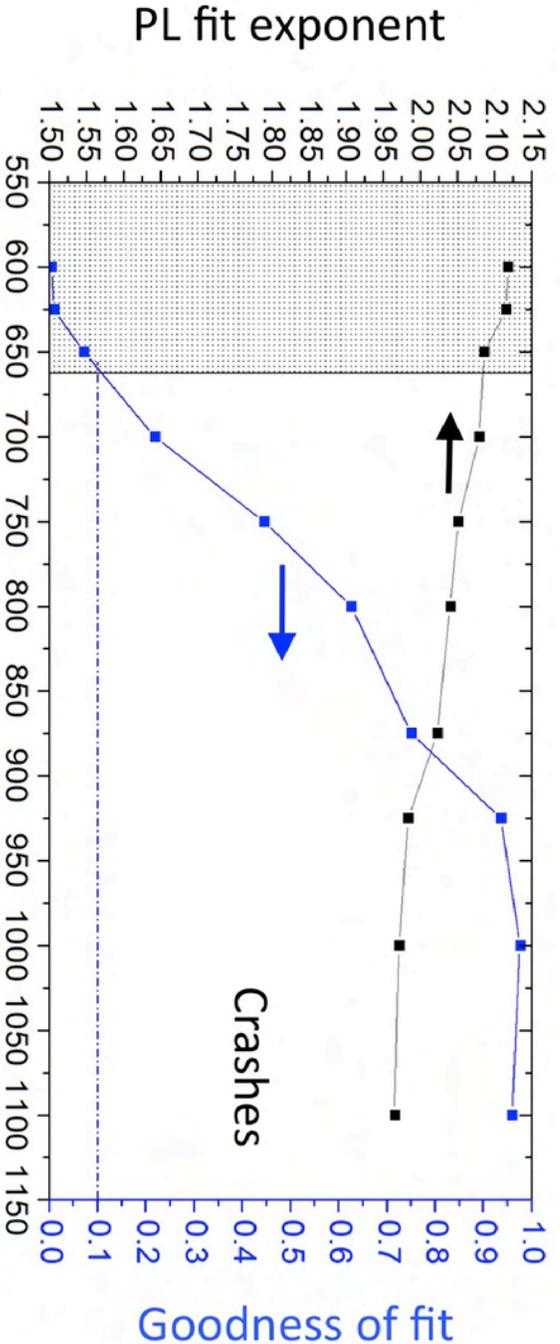
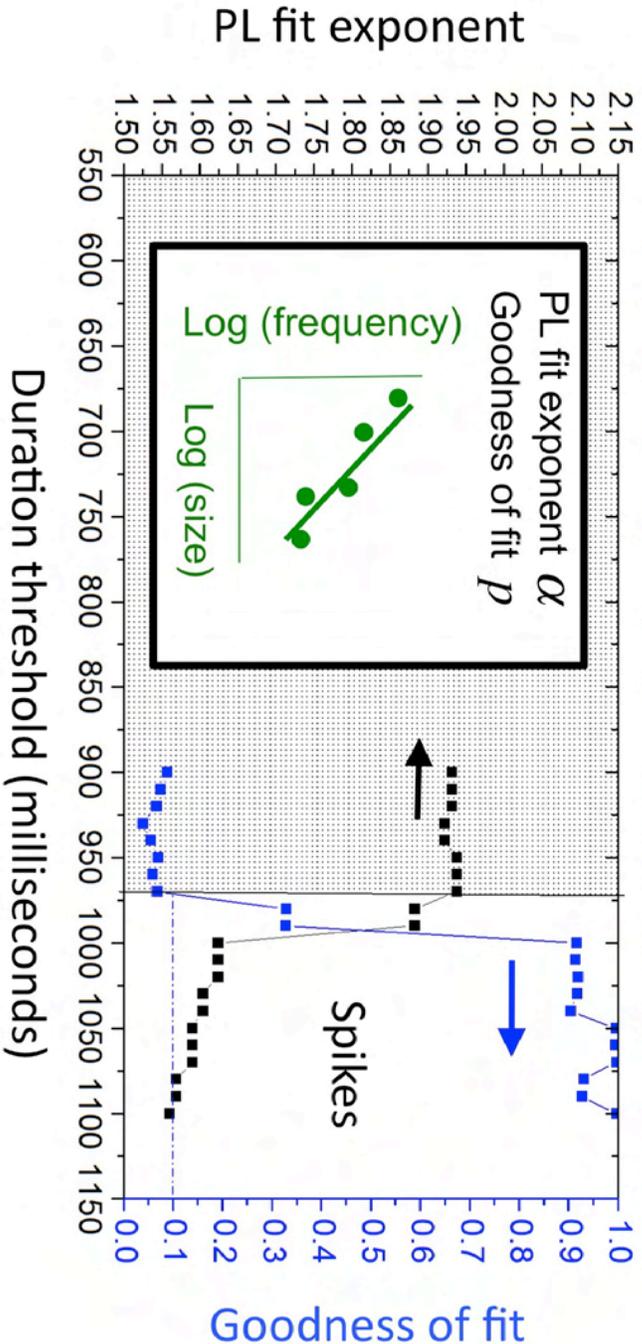

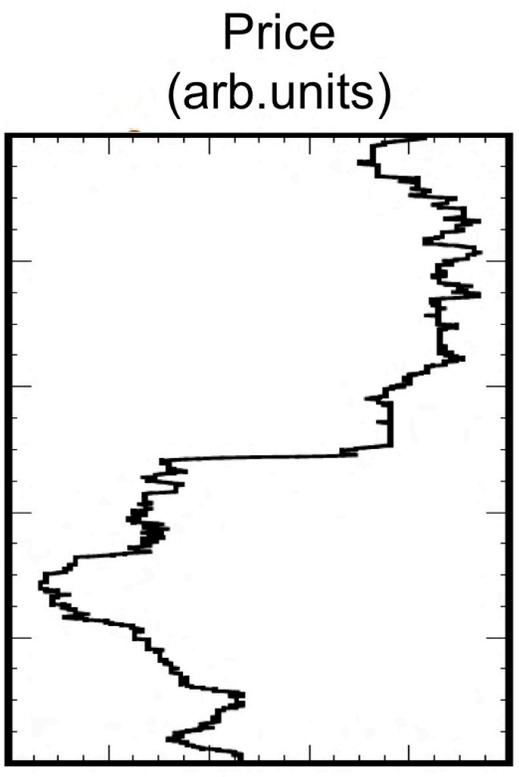

$\eta < 1$

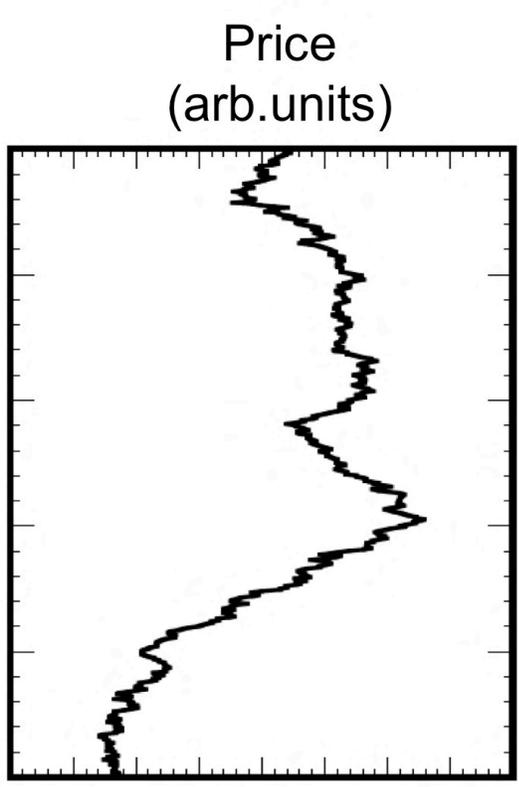

$\eta > 1$

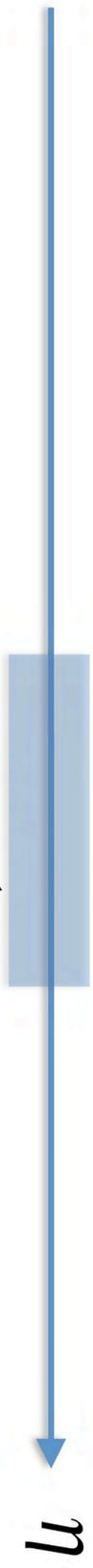

$\eta \approx 1$

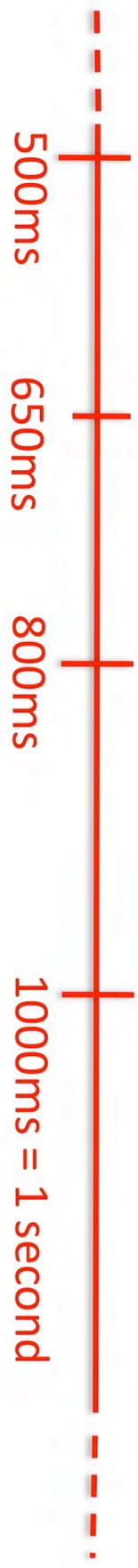

500ms

650ms

800ms

1000ms = 1 second

$\eta$

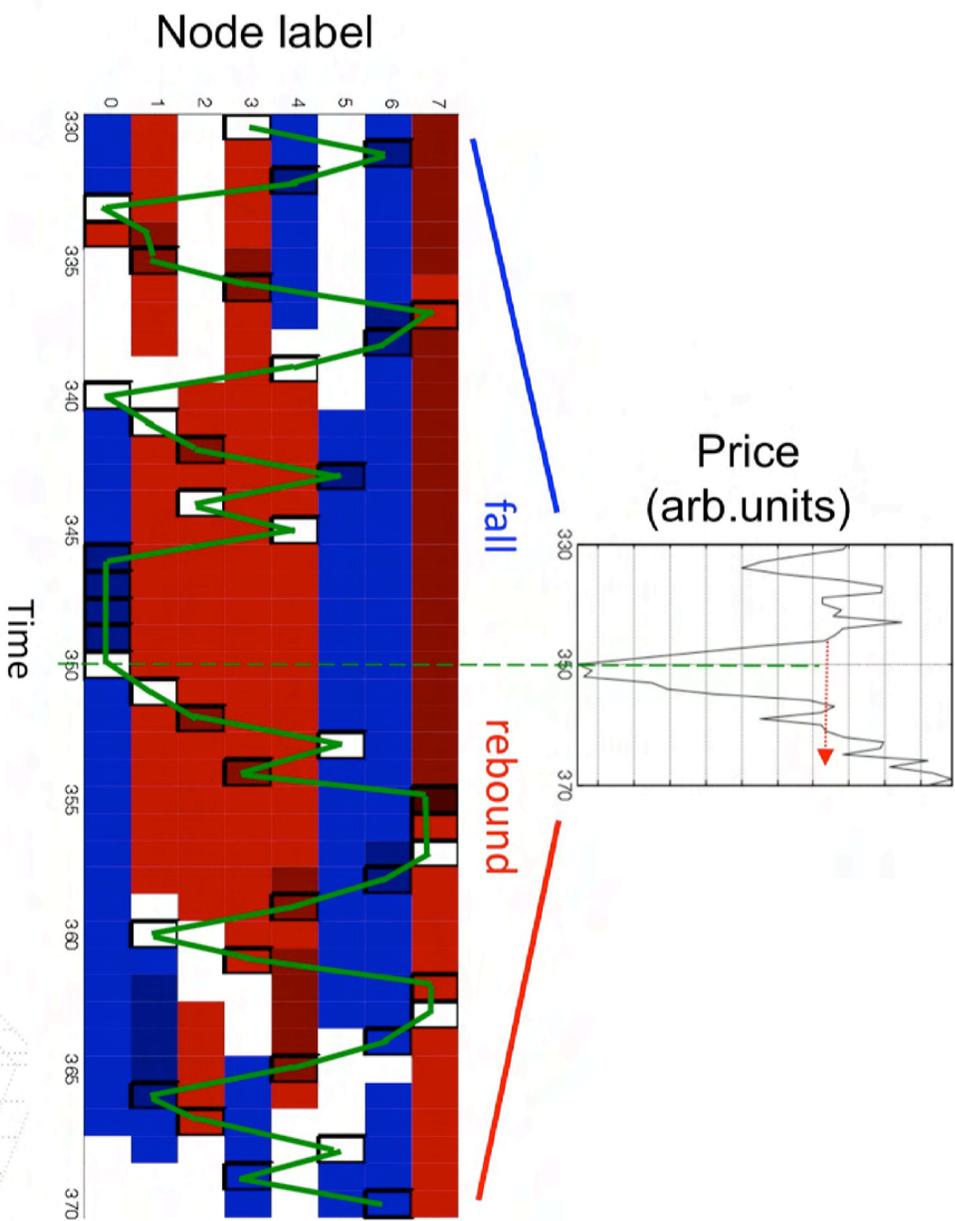
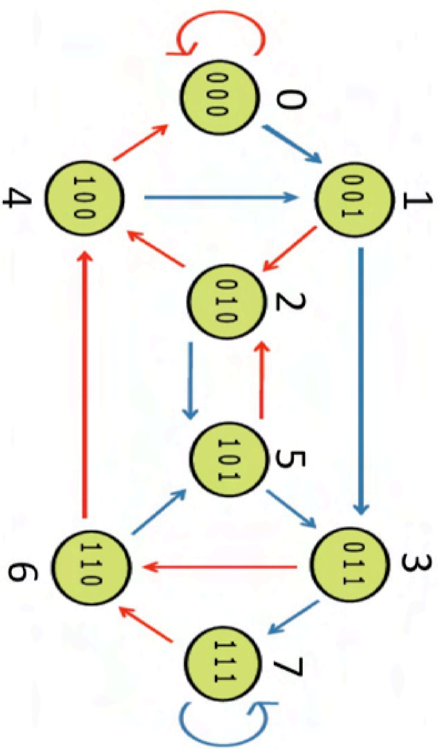
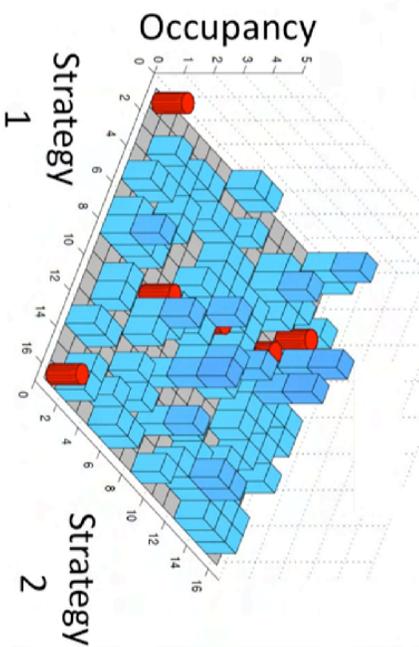